\documentclass[pra,superscriptaddress,nofootinbib,twocolumn,showpacs]{revtex4-1}
\usepackage{amsmath,graphicx,epsfig,color,amsfonts,amssymb,bbm}
\usepackage{hyperref}
\usepackage{amsthm}

\newcommand{\etal}{{\em et al.}}
\newcommand{\ket}[1]{|#1\rangle}
\newcommand{\bra}[1]{\langle#1|}
\newcommand{\braket}[1]{\left\langle #1 \right\rangle}
\newcommand{\one}{\mathbbm{1}}

\newcommand{\Conv}[1]{{\rm Conv}\!\left\{#1\!\right\}}
\newcommand{\vecP}{\vec{P}}

\newcommand{\Lc}[1]{{\mathcal L}_{#1}}
\newcommand{\Sv}[2]{{\mathcal S}_{#1/#2}}
\newcommand{\NS}[1]{\mathcal{NS}_{#1}}
\newcommand{\PTO}[1]{ \tilde{\mathcal T}_{#1}}
\newcommand{\Q}{\mathcal  Q}

\DeclareMathOperator{\tr}{tr}

\begin{document}

\title{Multipartite Nonlocality as a Resource and \\ Quantum Correlations Having Indefinite Causal Order}

\author{Florian Curchod}
\affiliation{Group of Applied Physics, University of Geneva, CH-1211 Geneva 4, Switzerland.}
\author{Yeong-Cherng~Liang}
\affiliation{Institute for Theoretical Physics, ETH Zurich, 8093 Zurich, Switzerland}
\affiliation{Group of Applied Physics, University of Geneva, CH-1211 Geneva 4, Switzerland.}
\author{Nicolas Gisin}
\affiliation{Group of Applied Physics, University of Geneva, CH-1211 Geneva 4, Switzerland.}

\date{\today}
\pacs{03.65.Ud, 03.67.Mn}

\begin{abstract}
The characterization of quantum correlations in terms of information-theoretic resource has been a fruitful approach to understand the power of quantum correlations as a resource.
While bipartite entanglement and Bell inequality violation in this setting have been extensively studied, relatively little is known about their multipartite counterpart. In this paper, we apply and adapt the recently proposed definitions of multipartite nonlocality [Phys. Rev. A {\bf 88}, 014102] to the three- and four-partite scenario to gain new insight on the resource aspect of multipartite nonlocal quantum correlations.
Specifically, we show that reproducing certain tripartite quantum correlations requires mixtures of classical resources --- be it the ability to change the groupings or the time orderings of measurements. Thus, when seen from the perspective of biseparable one-way classical signaling resources, certain tripartite quantum correlations do not admit a definite causal order. In the four-partite scenario, we obtain a superset description of the set of biseparable correlations which can be produced by allowing two groups of bipartite non-signaling resources. Quantum violation of the resulting Bell-like inequalities are investigated.  As a byproduct, we obtain some new examples of device-independent witnesses for genuine four-partite entanglement, and also device-independent witnesses that allows one to infer the structure of the underlying multipartite entanglement. 
\end{abstract}

\maketitle

\section{Introduction}
Quantum nonlocality, i.e., the fact that there exist quantum correlations which are stronger than that allowed by {\em any} locally-causal theory~\cite{Bell2004}, is one of the many counterintuitive features presented by quantum theory~\cite{Bell:1964}. Typically, this is demonstrated by the fact that some quantum correlations violate Bell inequalities~\cite{Brunner:RMP}, which are constraints that have to be satisfied by all locally-causal correlations. Empirically, the existence of non-locally-causal (hereafter abbreviated as nonlocal) correlations have also been verified --- modulo some logically possible but physically implausible loophole -- in a number of experiments (see, for instance, Refs.~\cite{Brunner:RMP,Experiments,Experiment3} and references therein).

With the advent of quantum information science, nonlocal quantum correlations have been dubbed a new role --- a  {\em resource}~\cite{Cleve2004,Barrett:PRA:2005,Brunner:NJP:2005} for information and communication tasks that are otherwise impossible in the classical world. A prime example of this is given by the task of secret key distribution, where nonlocal correlations have enabled the possibility to perform analysis in a paradigm where measurement devices need not be calibrated nor trusted --- the paradigm of {\em device-independent} quantum information processing~\cite{Scarani:DIQIP}. For instance, since entanglement is necessary to produce nonlocal correlations~\cite{Werner:PRA:1989}, the presence of bipartite or even genuine multipartite entanglement~\cite{Guhne:PhysRep} can be certified directly from measurement statistics through the violation of Bell inequalities~\cite{DIEW,Pal:DIEW, GUBI}. In fact, even entanglement quantification~\cite{Moroder:PRL:PPT} can be achieved in such a device-independent manner through Bell inequality violation (see also Ref.~\cite{Liang:PRA:2010}).

More concretely, nonlocal quantum correlations play a critical role in the security analysis of quantum cryptographic protocol where the devices are not trusted (see, for instance, Refs.~\cite{A.K.Ekert:PRL:1991,DIQKD} and references therein). They have also been used in randomness expansion~\cite{ BIV:Randomness}, in lower-bounding the underlying Hilbert space dimension~\cite{DimWitness,Moroder:PRL:PPT, Navascues:1308.3410}, in certifying that an entangling measurement has been performed~\cite{Rabello:PRL,TN,Navascues:1308.3410}, in certifying the teleportation of a qubit~\cite{Melvyn:2013}, in deducing the structure of the underlying multipartite entanglement~\cite{sharam,Moroder:PRL:PPT}, and  in self-testing of quantum devices~\cite{self-testing}.

While most of the tasks mentioned above concern a bipartite experimental setup, it is clear that genuine multipartite quantum nonlocality~\cite{svet87,SvetGeneralized,BBGL,chen,multinonlocality:measure,GUBI,Bancal:PRA:014102, Gallego:PRL:070401,Almeida:PRA:2010} --- the multipartite version of quantum nonlocality which relies on genuine multipartite entanglement --- must also be a resource in tasks where one does not want to rely on detailed characterization of measurement devices, or assumption on the underlying Hilbert space dimension. Somewhat surprisingly, not much on this is known in this regard (see Sec.~VI of Ref.~\cite{Brunner:RMP} for a recent review). In this work, we aim to gain further insight on this topic by focusing on the three- and four-partite scenario and comparing the extent to which classical one-way signaling resource, and post-quantum but non-signaling~\cite{PR:1994, Barrett:PRA:2005} resource can be used to reproduce quantum correlations.

The rest of this paper is organized as follows. In Sec.~\ref{Sec_Prelim}, we explain the various resources considered in this work; these include most of those defined in Ref.~\cite{Bancal:PRA:014102}, as well as an alternative that we propose in this work. Then, in Sec.~\ref{Sec_Mixtures}, we apply these definitions to the tripartite scenario and show that in order to reproduce certain tripartite quantum correlations, it is insufficient to have only the primitive form of these other resources, but one must also allow the possibility of mixing some of these resources. In Sec.~\ref{Sec_FourPartite}, we report progress on the characterization of biseparable correlations in the four-partite scenario. Specifically, we obtain a complete list of Bell-like inequalities that can be used to detect genuine tripartite (non-signaling) nonlocality in a four-partite scenario. 
Finally, we conclude with a summary and some further remarks in Sec.~\ref{Sec_Conclusion}.

\section{Preliminaries}\label{Sec_Prelim}

Let us begin by  providing a brief overview of the various probability distributions (correlations) needed for subsequent discussions. For a more detailed discussion on some of these definitions, see Refs.~\cite{Bancal:PRA:014102,Gallego:PRL:070401}. For simplicity, we focus predominantly on the tripartite scenario. Consider a Bell-type experiment where three parties Alice (A), Bob (B) and Charlie (C) are each allowed to perform, respectively, two alternative measurements (inputs) labeled by $x$, $y$, $z\in\{0,1\}$, and where each measurement gives two possible measurement outcomes (outputs) $a,b,c\in\{\pm1\}$. 

Following standard terminology, we say that a given tripartite conditional probability distribution $\vecP=\{P(abc|xyz)\}$ is {\em Bell-local} (henceforth abbreviated local) if it admits the decomposition~\cite{Brunner:RMP}:
\begin{equation}\label{Eq_local}
	P(abc|xyz) = \sum_\lambda q_\lambda P_\lambda(a|x)P_\lambda(b|y)P_\lambda(c|z)
\end{equation}
for all $x,y,z,a,b,c$, where $0\le q_\lambda\le 1$, $\sum_\lambda q_\lambda=1$ and $P_\lambda(a|x)$ is the conditional probability of getting outcome $a$ given the measurement setting $x$ and the hidden state $\lambda$; $P_\lambda(b|y)$ and $P_\lambda(c|z)$ are analogously defined. 
Operationally, the set of correlations satisfying Eq.~\eqref{Eq_local}, $\Lc{3}$, are those that can be produced classically using  {\em shared randomness} $\lambda$.

In contrast, the set of tripartite quantum correlations $\Q_3$ consists of correlations of the form of:
\begin{equation}\label{Eq_quantum}
	P(abc|xyz) = \tr (\rho\, M^x_a\otimes M^y_b\otimes M^z_c),
\end{equation}
where $\rho$ is a normalized quantum state, $\{M^x_a, M^y_b, M^z_c\}$ are local positive-operator-valued-measure
 elements describing Alice's, Bob's and Charlie's measurements respectively.
It is well-known~\cite{Bell:1964} that some quantum correlations are not Bell-local (nonlocal), in the sense that they cannot be written in the form of Eq.~\eqref{Eq_local}. However, as was first demonstrated by Svetlichny~\cite{svet87} in the tripartite scenario, certain quantum correlations can exhibit an even stronger form of nonlocality, in that they also cannot be decomposed in the following form:
\begin{equation}\label{Eq_bisep3}
	\begin{split}
		P(abc|xyz) &= \sum_\lambda q_\lambda P_\lambda(ab|xy)P_\lambda(c|z)\\
		&+\sum_\mu q_\mu P_\mu(ac|xz)P_\mu(b|y)\\
		&+\sum_\nu q_\nu P_\nu(bc|yz)P_\nu(a|x),
	\end{split}
\end{equation}
where $\sum\limits_{i \in\{ \lambda,\mu,\nu\}} q_i = 1$, $q_{i} \ge 0$ for all $i \in \{\lambda, \mu, \nu \}$ and $P_\lambda(ab|xy)$ etc. are {\em arbitrary} normalized bipartite conditional probability distributions. While correlations of the form~\eqref{Eq_bisep3} are not fully separable as in Eq.~\eqref{Eq_local}, they are  {\em biseparable} in the sense that they are formed by (mixtures of) correlations that are separable with respect to some bipartitions (groupings) of the parties.

Hereafter, we denote the set of such  biseparable correlations between a group of two parties and a group of single party by $\Sv{2}{1}$.  Operationally, correlations in $\Sv{2}{1}$ can be produced in the framework of a {\em nonlocal game}~\cite{Cleve2004} by allowing shared randomness between three parties and arbitrary classical communications within any subgroup of two parties.\footnote{Strictly, to recover the full set of such biseparable correlations, we must also allow the grouping to change from one round of the game to the next, and that there is no constraint in the measurement time ordering.} If some tripartite correlation $\vecP$ can be decomposed in the form of Eq.~\eqref{Eq_bisep3}, we say that $\vecP$ is $\Sv{2}{1}$-local, otherwise $\Sv{2}{1}$-nonlocal.

However, as has been pointed out in Refs.~\cite{Gallego:PRL:070401,Bancal:PRA:014102}, in a concrete physical scenario where measurements are performed according to some time ordering --- or equivalently where the inputs are given in some time-ordered manner --- the definition of biseparable correlations given in Eq.~\eqref{Eq_bisep3} may be physically unnatural and/or give rise to inconsistent prediction of the correlation $P(abc|xyz)$. As a result, two further (weaker) definitions were proposed in Refs.~\cite{Gallego:PRL:070401,Bancal:PRA:014102} to capture this stronger form of nonlocality that quantum correlations can exhibit.

The first of these imposes, on top of Eq.~\eqref{Eq_bisep3}, the requirement that all the multipartite probability distributions appearing in the decomposition also satisfy the non-signaling (NS) condition~\cite{PR:1994,Barrett:PRA:2005} at the level of the hidden state $\lambda$, i.e., 
\begin{subequations}\label{Eq_NS}
\begin{align}
	P_\lambda(a|x)&=\sum_b P_\lambda(ab|xy)\quad\forall\,\, a,x,y,\label{Eq_NSfromBToA}\\
	P_\lambda(b|y)&=\sum_a P_\lambda(ab|xy)\quad\forall\,\, b,x,y,\label{Eq_NSfromAToB}
\end{align}
\end{subequations}
and analogously for the other joint conditional probabilities $P_\mu(ac|xz)$ and $P_\nu(bc|yz)$.
Adopting the terminology from Ref.~\cite{Bancal:PRA:014102}, we shall refer to tripartite correlations satisfying Eq.~\eqref{Eq_bisep3} and all these NS conditions~\cite{PR:1994,Barrett:PRA:2005} as $\NS{2/1}$-local, otherwise $\NS{2/1}$-nonlocal (or three-way NS nonlocal).

The other definition proposed in Ref.~\cite{Bancal:PRA:014102} allows one-way signaling resources but in order to exploit these signaling resources in a physically meaningful manner, the time ordering of the measurements matters. Specifically, the definition of Ref.~\cite{Bancal:PRA:014102} consists of imposing:
\begin{equation}\label{Eq_T2_1}
	\begin{split}
		P(abc|xyz) &= \sum_\lambda q_\lambda P^{T_{AB}}_\lambda(ab|xy)P_\lambda(c|z)\\
		&+\sum_\mu q_\mu P^{T_{AC}}_\mu(ac|xz)P_\mu(b|y)\\
		&+\sum_\nu q_\nu P^{T_{BC}}_\nu(bc|yz)P_\nu(a|x),
	\end{split}
\end{equation}
{\em for all}\, possible time orderings of the measurement events, i.e., if Alice measures first, followed by Bob, followed by Charlie, which we denote by $T_{ABC}=A<B<C$, or if Alice measures first, followed by Charlie, followed by Bob, $T_{ABC}=A<C<B$, etc. In the above equation, $P^{T_{AB}}_\lambda(ab|xy)$ denotes, for the time ordering $T_{AB}$, the conditional probability distribution of getting outcome $a$, $b$ given the hidden state $\lambda$ and the measurement inputs $x$, $y$. For instance if $T_{AB}=A<B$,  one-way signaling is allowed from Alice to Bob, i.e., Eq.~\eqref{Eq_NSfromBToA} is still required to hold but Eq.~\eqref{Eq_NSfromAToB} may be violated since $b$ may depend explicitly on $y,x$ and $a$ in this case. 

Inspired by the above definition, we consider in this paper correlations that can be decomposed in the following form:
\begin{equation}\label{Eq_PTO}
\begin{split}
P(abc|xyz) = \sum\limits_{\lambda_1} q_{\lambda_1}  P_{\lambda_1}^{A<B}(ab|xy)  P_{\lambda_1}(c|z) \\
		+ \sum\limits_{\lambda_2} q_{\lambda_2}  P_{\lambda_2}^{B<A}(ab|xy)  P_{\lambda_2}(c|z) \\
		+ \sum\limits_{\mu_1} q_{\mu_1}  P_{\mu_1}^{A<C}(ac|xz)  P_{\mu_1}(b|y) \\
		+ \sum\limits_{\mu_2} q_{\mu_2}  P_{\mu_2}^{C<A}(ac|xz)  P_{\mu_2}(b|y) \\
		+ \sum\limits_{\nu_1} q_{\nu_1}  P_{\nu_1}^{B<C}(bc|yz)  P_{\nu_1}(a|x) \\
		+ \sum\limits_{\nu_2} q_{\nu_2}  P_{\nu_2}^{C<B}(bc|yz)  P_{\nu_2}(a|x),
\end{split}
\end{equation}
where 
\begin{equation}
	\sum\limits_{i \in \{\lambda_k,\mu_k,\nu_k\}_{k=1,2}} q_i = 1, \quad q_i\ge0\quad\forall\,\, {i = \{\lambda_k,\mu_k,\nu_k\}_{k=1,2}}, 
\end{equation}
and the bipartite distributions $P_{\lambda_1}^{A<B}(ab|xy)$ etc. are defined as in the last paragraph.
Denoting this set of correlations by $\PTO{2/1}$, we see that $\PTO{2/1}$ is the set of correlations that can be achieved by classical systems if we allow shared randomness between all parties, and also one-way signaling resources for some given time ordering and where both {\em the  grouping of parties} and  {\em the time ordering} may change from one run to the next. Hence, the time ordering resources described by Eq.~\eqref{Eq_PTO} differ from that considered in Refs.~\cite{Gallego:PRL:070401,Bancal:PRA:014102}. More explicitly, as discussed in Sec.~\ref{Sec_Mixtures}, $\PTO{2/1}$ can be seen as the {\em convex hull} \footnote{The convex hull of a set $\mathcal{A}$ is the set formed by all possible convex combinations of points in $\mathcal{A}$.}  of all biseparable one-way signaling correlations admitting different time orderings, whereas the $K_2$-local set defined in Ref.~\cite{Bancal:PRA:014102} (also called the known-sequence model in Ref.~\cite{Pironio:JPA:065303}) amounts, instead,  to their {\em intersection} and hence being a strict subset of $\PTO{2/1}$; the $T_2$-local set defined in Ref.~\cite{Bancal:PRA:014102} is an even smaller subset of $\PTO{2/1}$.

For any finite set of measurement choices and outcomes, all but the quantum sets defined above are {\em convex polytopes}, i.e., convex sets having  finite number of extreme points and hence can be equivalently described by a finite set of Bell (-like) inequalities.\footnote{The fact that $\Lc{n}$ is a polytope was discussed in Ref.~\cite{Pitowsky:Book}. For the other sets, we refer the reader to Ref.~\cite{Bancal:PRA:014102} for a detailed discussion.} 
The set of quantum correlations, though convex~\cite{Pitowsky:Book}, generally does not have a simple characterization in terms of its extreme points. The same holds true for the set of biseparable quantum correlations $\Q_{2/1}$~\cite{DIEW}, i.e., those obtained by performing local measurements on a tripartite quantum state that is biseparable~\cite{Guhne:PhysRep}.

Evidently, the terminologies introduced above admit natural generalization to the $n$-partite scenario. 
Moreover, for the benefit of later discussion, it is worth noting the following inclusion relations:
\begin{equation}\label{Eq_Sets}
\begin{split}
	\Lc{n}\subset \Q_{k/n-k} \subset \NS{k/n-k}  \subset \PTO{k/n-k} \subset \Sv{k}{n-k}
\end{split}
\end{equation}
and that that even though $\Q_{k/n-k}\subset \Q_n$ for all $k< n$, 
the sets $\NS{k/n-k}$, $\PTO{k/n-k}$,  $\Sv{k}{n-k}$ generally cannot be compared with $\Q_n$.

\section{Reproducing certain quantum correlations requires mixtures}
\label{Sec_Mixtures}

Now, let us apply the definitions provided in the last section to gain some insights on when it is possible, and what is required, to reproduce certain tripartite quantum correlations using the classical/ post-quantum resources mentioned above.

\subsection{Necessity of mixing different time-ordered resources}
\label{Sec:TimeOrdered}

The definition of $\PTO{2/1}$, as can be seen in Eq.~\eqref{Eq_PTO}, involves mixtures of different bipartitions and different time orderings, for instance, having $A$ and $B$ in the same group and $A$ measuring before $B$ with probability $\sum_{\lambda_1} q_{\lambda_1}$ etc. One can, of course, also analyze subsets of $\PTO{2/1}$ corresponding to a fixed bipartition and fixed time ordering. 
For example, let us denote by $\PTO{2/1}^{(A<B)}$ the set of tripartite correlations  admitting the decomposition
\begin{equation}\label{Eq_bisep3_T12}
	P(abc|xyz) = \sum\limits_{\lambda_1} q_{\lambda_1}  P_{\lambda_1}^{A<B}(ab|xy)  P_{\lambda_1}(c|z)
\end{equation}
with $ \sum\limits_{\lambda_1} q_{\lambda_1}=1$,  and analogously for $\NS{2/1}^{(A<C)}$, $\NS{2/1}^{(B<C)}$ etc. Likewise, let us denote by $\PTO{2/1}^{(A<B<C)}$ the set of correlations that can be obtained by having a {\em definite time ordering} $T_{ABC}=A<B<C$ but with the possibility of mixing different bipartitions, i.e., 
\begin{equation}\label{Eq_bisep3_T123}
\begin{split}
P(abc|xyz) = \sum\limits_{\lambda_1} q_{\lambda_1}  P_{\lambda_1}^{A<B}(ab|xy)  P_{\lambda_1}(c|z) \\
		+ \sum\limits_{\mu_1} q_{\mu_1}  P_{\mu_1}^{A<C}(ac|xz)  P_{\mu_1}(b|y) \\
		+ \sum\limits_{\nu_1} q_{\nu_1}  P_{\nu_1}^{B<C}(bc|yz)  P_{\nu_1}(a|x).
\end{split}
\end{equation}
where $\sum_{i\in\{\lambda_1, \mu_1,\nu_1\}}  q_i=1$; the sets correspond to other fixed time orderings can be similarly defined. Clearly, the set  $\PTO{2/1}$ as defined in Eq.~\eqref{Eq_PTO} can also be written as the {\em convex hull} of these constituent polytopes, i.e., 
\begin{equation}\label{Eq_PTO_CvHull}
\begin{split}
	\PTO{2/1}={\rm Conv}\Big\{&\PTO{2/1}^{(A<B)},\PTO{2/1}^{(A<C)},\PTO{2/1}^{(B<C)},\\
				&\PTO{2/1}^{(B<A)},\PTO{2/1}^{(C<A)},\PTO{2/1}^{(C<B)}\Big\},\\
			    ={\rm Conv}\Big\{&\PTO{2/1}^{(A<B<C)},\PTO{2/1}^{(C<B<A)}\Big\}.
\end{split}
\end{equation}

Now, one may ask whether the mixtures involved in Eq.~\eqref{Eq_PTO} are {\em necessary} in order to reproduce all quantum correlations in $\PTO{2/1}$, i.e., all quantum correlations that are biseparable with respect to these one-way signaling resources. For instance, does there exist
$\vecP\in\Q_3$ such that $\vecP\in\PTO{2/1}$ but $\vecP\not\in \PTO{2/1}^{(A<B)}\cup\PTO{2/1}^{(A<C)}\cup\PTO{2/1}^{(B<C)}\PTO{2/1}^{(B<A)}\cup\PTO{2/1}^{(C<A)}\cup\PTO{2/1}^{(C<B)}$, i.e., $\vecP$ {\em cannot} be obtained by allowing a {\em fixed} subset of the parties sharing one-way signaling resources and obeying definite time ordering? With some thought, it is easy to see that this can be trivially achieved by performing local measurements on some (in)coherent superposition of tripartite quantum states that have orthogonal (local)  supports. But what if we restrict our attention to local measurement on a three-qubit state, i.e., one where this trivial encoding of classical mixture into the quantum state is not possible? In the following, we provide an affirmative answer to this  question.

To this end, let us recall the $W$-state~\cite{Dur:PRA:062314} (written in the computational basis):
\begin{equation}\label{Eq_WState}
	\ket{W} = \frac{1}{\sqrt{3}} (\ket{100}+\ket{010}+\ket{001})
\end{equation}
and the following choice of $\pm1$-outcome Hermitian observables~\cite{Wurflinger:PRA:032117}
\begin{gather}
A_0 = \cos\alpha \, \sigma_z + \sin\alpha \, \sigma_x,\quad
A_1 = \cos\alpha \, \sigma_z - \sin\alpha \, \sigma_x,\nonumber \\ 
B_0 = -\sigma_z,\quad B_1 = \cos\beta \, \sigma_z + \sin\beta \, \sigma_x,
\label{Eq_Obs_WState}\\ 
C_0 = -\sigma_z, \quad C_1 = \cos\beta \, \sigma_z - \sin\beta \, \sigma_x \nonumber,
\end{gather}
where $\alpha = 3.6241$, $\beta = 2.0221$, $\sigma_x,\sigma_y,\sigma_z$ are the Pauli matrices and the measurement setting is labeled by the subscript attached to each observable. 

Let us denote by $\vecP_W$ the tripartite probability distributions obtained by applying the measurements associated with Eq.~\eqref{Eq_Obs_WState} to $\ket{W}$. It is known~\cite{Wurflinger:PRA:032117} that $\vecP_W\not\in\Sv{2}{1}$ and is thus $\Sv{2}{1}$-nonlocal. Consider now the mixture of $\vecP_W$ with the uniform probability distribution $\vecP_0$:
\begin{equation}\label{Eq_Pv_dfn}
	\vecP_W'(v)= v \, \vecP_W + (1-v) \vecP_0,\quad 0\le v\le 1.
\end{equation}
It is easy to see that since  $\vecP_0\in\Lc{3}$, it is also a member of all the sub-polytopes of $\PTO{2/1}$. Hence, for each of these polytopes, there exists a maximal value of $v=v_{\mbox{\tiny max}}$ --- often referred to as the  {\em visibility} --- such that $\vecP_W'(v_{\mbox{\tiny max}})$ is still a member of the polytope of interest. Since one can  enumerate all the extreme points of these individual polytopes, the corresponding $v_{\mbox{\tiny max}}$ can be determined efficiently by solving a linear program. The results are summarized in Table~\ref{Tbl:TimeOrderedPolytope}.

\begin{table}[h!]
        \begin{tabular}{|c|c|}\hline
        		Set(s)  &   $v_{\mbox{\tiny max}}$  \\ \hline
		$\Sv{2}{1}$ &  0.9548 \\ 
$\PTO{2/1}$ & 0.9339 \\ 
$\PTO{2/1}^{B<C<A}$ & 0.9138 \\
$\PTO{2/1}^{A<B<C}$, $\PTO{2/1}^{A<C<B}$, $\PTO{2/1}^{B<A<C}$,
$\PTO{2/1}^{C<A<B}$, $\PTO{2/1}^{C<B<A}$ & 0.8931 \\
$\Conv{\PTO{2/1}^{(B<C)},\PTO{2/1}^{(C<B)}}$, & 0.8420 \\
$\Conv{\PTO{2/1}^{(A<B)},\PTO{2/1}^{(B<A)}}$,
$\Conv{\PTO{2/1}^{(A<C)},\PTO{2/1}^{(C<A)}}$ & 0.8318 \\
$\PTO{2/1}^{(A<B)}$, $\PTO{2/1}^{(A<C)}$ & 0.8212 \\
$\PTO{2/1}^{(B<A)}$, $\PTO{2/1}^{(C<A)}$, 
$\PTO{2/1}^{(B<C)}$, $\PTO{2/1}^{(C<B)}$ & 0.7120 \\ 
		$\Lc{3}$ &  0.7120 \\\hline
	\end{tabular}
	\caption{\label{Tbl:TimeOrderedPolytope} Maximal value of $v$ for the correlation $\vecP_W'(v)$ defined in Eq.~\eqref{Eq_Pv_dfn} to remain inside the various time-ordered convex sets.}
\end{table}

Clearly,  from Table~\ref{Tbl:TimeOrderedPolytope}, we see that some of the quantum correlations $\vecP_W'(v)$ can only be reproduced if we allow the parties in the subgroup to change from one run to the next. For instance, if we only allow Alice and Bob to form a subgroup, and one-way signaling from Bob to Alice, then we cannot reproduce $\vecP_W'(v)$ any more than we can do using resources in $\Lc{3}$.
Likewise, even if we allow the direction of one-way signaling to change, as long as other grouping of parties are not allowed, we can at best reproduce $\vecP_W(v)$ for $v\le 0.8420$. Note also that the difference in these critical values of $v$ indicate that there are $\vecP_W'(v)$ that can be reproduced {\em only if} we allow changes in the time ordering, or equivalently mixtures of these one-way signaling resources. Moreover, this is true even if we allow arbitrary mixtures of the groupings of parties; mixtures of time ordering still offers an advantage in terms of the range of $v$ whereby $\vecP_W'(v)$ can be reproduced in this classical manner. 

The above observation implies that, as seen from the perspective of these classical,  one-way signaling resources, some quantum correlations do not have a well-defined {\em causal order},\footnote{Here, we borrow the terminology from Ref.~\cite{Oreshkov:NatComm:1092} that was used in a different context. Instead, the lack of causal order as we define here, is analogous to the lack of definite grouping/ bipartitioning for certain biseparable quantum states in the studies of quantum entanglement.}  i.e., reproducing them requires resources corresponding to different time orderings. Note that this does not imply that quantum correlations cannot be reproduced using one-way signaling resources. However, any plausible one-way signaling resource that reproduces all quantum correlations must lead to the possibility of superluminal communication~\cite{Bancal:NatPhys,Barnea:2013}. Bearing this in mind, it is natural to ask if the necessity to mix resources corresponding to different bipartitions is an artifact of choosing (one-way) signaling resources, or whether we also observe the same effect using NS resources. To answer this question, let us next investigate the corresponding scenario when the parties in each subgroup are only allowed to shared NS resources.

\subsection{Necessity of mixing different non-signaling resources}
\label{Sec:MixPart}

In analogy with the previous section, let us denote by $\NS{2/1}^{(A,B)}$ the set of tripartite correlations  admitting the decomposition
\begin{equation}\label{Eq_bisep3_sub12}
	P(abc|xyz) = \sum_\lambda q_\lambda P_\lambda(ab|xy)P_\lambda(c|z)
\end{equation}
with $\sum_\lambda q_\lambda=1$ while satisfying Eq.~\eqref{Eq_NS},  and analogously for $\NS{2/1}^{(A,C)}$ and $\NS{2/1}^{(B,C)}$. Evidently, the set  $\NS{2/1}$ can also be written as the convex hull of the three sub-polytopes, i.e., 
\begin{equation}\label{Eq_NS_CvHull}
	\NS{2/1}=\Conv{\NS{2/1}^{(A,B)},\NS{2/1}^{(A,C)},\NS{2/1}^{(B,C)}}.
\end{equation}
These sub-polytopes  are basically the sets of correlations where only the parties labeled in the superscripts are allowed to share arbitrary NS resources, and that no mixtures between different groupings of the parties are allowed. Again, if we consider the quantum correlations defined in Eqs.~\eqref{Eq_WState}-\eqref{Eq_Pv_dfn}, we can determine the critical value $v$ beyond which $\vecP_W'(v)$ is outside the various polytopes; the results are summarized in Table~\ref{Tbl_NS_Visibilities}.

\begin{table}[h!]
        \begin{tabular}{|c|c|}\hline
		Set(s)  &   $v_{\mbox{\tiny max}}$  \\ \hline
		$\NS{2/1}$ &  0.8477 \\
		$\Conv{\NS{2/1}^{(A,B)},\NS{2/1}^{(A,C)}}$ &   0.8212 \\
		$\Conv{\NS{2/1}^{(A,B)},\NS{2/1}^{(B,C)}}$, $\Conv{\NS{2/1}^{(A,C)},\NS{2/1}^{(B,C)}}$&  0.7120 \\
		$\NS{2/1}^{(A,B)}$ , $\NS{2/1}^{(A,C)}$ , $\NS{2/1}^{(B,C)}$ &  0.7120 \\
		$\Lc{3}$ &  0.7120 \\\hline
	\end{tabular}
	\caption{\label{Tbl_NS_Visibilities} Maximal value of $v$ for the correlation $\vecP'_W(v)$ defined in Eqs.~\eqref{Eq_WState}-\eqref{Eq_Pv_dfn} to remain as a member of the various convex sets.}
\end{table}

From Table~\ref{Tbl_NS_Visibilities}, it is clear that in Eq.~\eqref{Eq_Pv_dfn}, different weights of $\vecP_0$ are required for $P(v)$ to remain as a member of $\NS{2/1}$ or its ``constituents" $\NS{2/1}^{(A,B)}$ etc. In fact, for $v>0.8212$, it is even insufficient to allow the mixtures of two different groupings, but we must necessarily allow all possible bipartitions in the mixtures in order to reproduce $P_W'(v)$.

\subsection{Common boundary of polytopes}

Evidently, the necessity of mixtures in the above discussion should not be taken for granted as the generic behavior of what is required to reproduce quantum correlations. For instance, if we had instead started with some other three-qubit quantum correlations $\vecP$ instead of $\vecP_W$, it could well be that the critical value $v$ may coincide for the case with or without mixtures of groupings/ time ordering. 
Indeed, an explicit example of this kind is given by measuring the following observables\footnote{Note that there is a misprint in  the specification of $X_3$ and $X_3'$ in Ref.~\cite{BBGL}: instead of $X_3=\sigma_x$, $X_3'=-\sigma_y$, it should have been $X_3=-\sigma_y$ and $X_3=\sigma_x$. }~\cite{BBGL}
\begin{gather}
A_0 = \sigma_x,\quad A_1 = \sigma_y ,\nonumber \\ 
B_0 = \frac{1}{\sqrt{2}} (\sigma_x - \sigma_y) ,\quad B_1 =\frac{1}{\sqrt{2}} (\sigma_x + \sigma_y),\\ 
C_0 = -\sigma_y , \quad C_1 = \sigma_x,\nonumber 
\end{gather}
on the tripartite Greenberger-Horne-Zeilinger (GHZ) state~\cite{GHZ}
\begin{equation}
	\ket{\rm GHZ_3} = \frac{1}{\sqrt{2}}  (\ket{000}+\ket{111}),
\end{equation}
which violates the Svetlichny inequality~\cite{svet87,BBGL}
\begin{equation}
 -\braket{A_0 B_0 C_0}+ \braket{A_0 B_0 C_1}+\braket{A_0 B_1 C_1}-\braket{A_1 B_1 C_1} 
+\circlearrowright\,\, \le 4,\label{Ineq_Svet}
\end{equation}
up to a value of $4\sqrt{2}$; here and below, we use $\circlearrowright$ to denote additional terms which must be added to ensure that the inequality is invariant under arbitrary permutation of parties.

If we again mix the resulting correlations $\vecP_G$ with the uniform probability distributions, it can be verified that if, and only if,  $v\le\tfrac{1}{\sqrt{2}}$, the resulting mixture $\vecP_G'(v)=v \, \vecP_G + (1-v) \vecP_0$ is always inside all the polytopes considered in Sec.~\ref{Sec:TimeOrdered} and Sec.~\ref{Sec:MixPart}. This means that $\vecP_G(v)$ crosses all these polytopes, including the local polytope,  at the same critical value of $v$. Thus, although the various convex sets defined above are clearly distinct, they also share some nontrivial common boundary.

\section{Towards a characterization of the set of biseparable correlations in the four-partite scenario}
\label{Sec_FourPartite}

While the three-partite scenario with binary inputs and outputs are relatively well-studied\footnote{In terms of facet-defining inequalities, the sets $\Lc{3}$ and $\NS{2/1}$ were fully characterized, respectively, in Ref.~\cite{Sliwa} and in Ref.~\cite{Bancal:PRA:014102}. A numerical method for characterizing the set $\Q_{2/1}$ was also given in Ref.~\cite{DIEW}.}, very little --- in terms of Bell-like inequalities --- is known about the various sets of correlations in the four-partite scenario (see, however, Refs.~\cite{WW:Complete:npartiteBI,ZB:Complete:npartiteBI,Bancal:JPA:385303}). For the set of biseparable correlations, this characterization is further complicated by the fact that there are now two different types of groupings, i.e., those characterized by a subgroup of one vs a subgroup of three, 
\begin{equation}\label{Eq_bisep4_3vs1}
\begin{split}
P(abcd|xyzw)  
&=\sum_\lambda q_\lambda P_\lambda(abc|xyz)P_\lambda(d|w)\\
&+\sum_\mu q_\mu P_\mu(abd|xyw)P_\mu(c|z)\\
&+\sum_\nu q_\nu P_\nu(acd|xzw)P_\nu(b|y),\\
&+\sum_\theta q_\theta P_\theta(bcd|yzw)P_\theta(a|x),
\end{split}
\end{equation}
where $\sum\limits_{i = \lambda,\mu,\nu,\theta} q_i = 1$ and $q_{i} \ge 0$ for all $i \in \{\lambda, \mu, \nu,\theta \}$, and those characterized by two subgroups of two,
\begin{equation}\label{Eq_bisep4_2vs2}
\begin{split}
P(abcd|xyzw) &= \sum_\lambda q_\lambda P_\lambda(ab|xy)P_\lambda(cd|zw)\\
&+\sum_\mu q_\mu P_\mu(ac|xz)P_\mu(bd|yw)\\
&+\sum_\nu q_\nu P_\nu(ad|xw)P_\nu(bc|yz),
\end{split}
\end{equation}
where $\sum\limits_{i = \lambda,\mu,\nu} q_i = 1$ and $q_{i} \ge 0$ for all $i \in \{\lambda, \mu, \nu \}$. Here, we have labeled the fourth party's input (output) by $w$ ($d$). 

Hereafter, we focus on  the simplest set of correlations satisfying the above decomposition conditions, namely, those  where the bipartite and tripartite terms satisfy the analogous NS conditions [cf Eq.~\eqref{Eq_NS}]. Using the terminology of Sec.~\ref{Sec_Prelim}, these convex sets are thus, respectively, the $\NS{3/1}$ and $\NS{2/2}$ polytope. In contrast with the tripartite scenario where three-way NS nonlocality can be identified with being $\NS{2/1}$-nonlocal, a given correlation is said to exhibit four-way NS-nonlocality only if it is outside the convex hull of  $\NS{3/1}$ and $\NS{2/2}$.
Building on the characterization of $\NS{3}$ polytope achieved in Ref.~\cite{Pironio:JPA:065303}, one can enumerate all the 860,160 extreme points~\cite{Florian:Thesis} of $\NS{3/1}$ in $\mathbb{R}^{80}$.
Likewise, using the fact that with binary inputs and outputs, there is only one non-trivial family of extremal NS probability distributions --- the Popescu-Rohrlich (PR) box~\cite{PR:1994}, one can enumerate all the 1,216 extreme points~\cite{Florian:Thesis} of $\NS{2/2}$ in $\mathbb{R}^{80}$. However, even in this latter case, it seems formidable to use a standard polyhedron representation software such as PORTA~\cite{PORTA} to determine the complete set of facet-defining Bell-like inequalities of $\NS{2/2}$ within a reasonable amount of time.

\subsection{The Symmetrized $\NS{2/2}$ Polytope}

As a compromise, we employed the trick of Ref.~\cite{Bancal:JPA:385303} and solved, instead for the symmetrized $\NS{2/2}$ polytope, i.e., the projection of the $\NS{2/2}$ polytope onto the subspace spanned by all correlations that are invariant under arbitrary permutation of parties.\footnote{This polytope can be described by 116 extreme points in $\mathbb{R}^{14}$~\cite{Florian:Thesis}.} In other words, we obtained a minimal set of Bell-like inequalities having the same symmetry and which are satisfied by all correlations in $\NS{2/2}$. Henceforth, we refer to these Bell-like inequalities as the symmetrical $\NS{2/2}$ inequalities. Their permutational invariance means that if
\begin{equation}
	\sum_{a,b,c,d,x,y,z,w} \gamma^{xyzw}_{abcd} P(abcd|xyzw) \stackrel{\NS{2/2}}{\le} \beta,
\end{equation}
is one such inequality that is satisfied by all correlations of $\NS{2/2}$, where $\gamma^{xyzw}_{abcd}\in\mathbb{R}$ then $\gamma^{xyzw}_{abcd}=\pi(\gamma^{yxzw}_{bacd})=\ldots$ etc. for all possible permutations $\pi$ of the input-output indices associated with the parties.

There are in total 180,006 of these Bell-like inequalities, and together they provide a minimal but complete description of the symmetrized $\NS{2/2}$ polytope. Exploiting the freedom in labeling the input, output and parties, these inequalities can be further categorized into 23,306 inequivalent families.\footnote{Two inequalities are said to be equivalent if we can obtain one from the other through relabeling of the indices of the input and/or output and/or party~\cite{WW:Complete:npartiteBI}.} Among them, 13,479 are also facet-defining for the original (non-symmetrized) $\NS{2/2}$ polytope. For a representative of the complete list of all these 23,306 inequalities and some of their properties,  we refer the reader to Ref.~\cite{URL}. 

Let us now highlight some interesting features of this polytope. Firstly, as with the facet-defining inequalities of $\NS{2/1}$ obtained in Ref.~\cite{Bancal:PRA:014102}, all these symmetrical $\NS{2/2}$ inequalities  are also saturated by some extreme point(s) of $\Lc{4}$. However, except for the positivity facet --- an inequality which dictates that probability has to be non-negative --- none of these inequalities corresponds to a facet of $\Lc{4}$. The above observation nevertheless implies that a  violation of any of these inequalities immediately implies not only Bell-nonlocality, but also the stronger form of  three-way NS-nonlocality, i.e., any possible decomposition of the Bell-inequality-violating probability distributions must involve genuine tripartite terms like $P(abc|xyz,\lambda)$.   

In addition, it is worth noting that except for 9 inequivalent families of the symmetrical $\NS{2/2}$ inequalities, all the rest of them can be violated by some extreme point(s) of  the $\NS{3/1}$ polytope. Note, however, that the set $\NS{3/1}$ and $\NS{2/2}$ are not comparable. In particular, the correlations corresponding to two subgroups each having access to a PR-box cannot be decomposed in the form of Eq.~\eqref{Eq_bisep4_3vs1}.

\subsection{Quantum Violation}
 
Clearly, since our goal is to understand the resource aspect of multipartite nonlocal quantum correlations, it is of interest to understand the possibility to violate these inequalities using quantum correlations. In this regard, we note that except for the family of positivity facets, and 10 other families of $\NS{2/2}$ inequalities, all other inequalities defining the symmetrical $\NS{2/2}$ polytope can be violated by some 4-qubit entangled states. Geometrically,  this means that in the abstract space of correlations, there are many directions (as defined by the coefficients of some Bell-like inequalities) where an $\NS{2/2}$ resource is insufficient to reproduce all quantum correlations.
In particular, the direction corresponding to the following inequality --- which also defines a facet of the (non-symmetrized) $\NS{2/2}$ polytope --- was found to give the best visibility and hence  resistance to white noise:
\begin{align}
I_{\rm opt} = &\braket{A_0 B_0 C_0 D_0}+ 2\braket{A_0 B_0 C_1 D_1}-8\braket{A_1 B_1 C_1 D_1} \nonumber\\
&- 3\braket{A_0 B_0 C_0}+2 \braket{A_0 B_1 C_1}- \braket{A_0 B_0 }+2\braket{A_1 B_1 }\nonumber\\
&-\braket{A_0 }+\circlearrowright \quad\stackrel{\NS{2/2}}{\le} 19,\label{Ineq_OptimalVis}
\end{align}
where the correlators --- the expectation values of the product of outcomes --- are defined in terms of the joint (or marginal) probability distributions as 
\begin{equation}
\begin{split}
	\braket{A_x B_y C_z D_w}&=\sum_{a,b,c,d=\pm1} a\,b\,c\,d\, P(abcd|xyzw),\\
	\braket{A_x B_y C_z}&=\sum_{a,b,c=\pm1} a\,b\,c\, P(abc|xyz),\\
	\braket{A_x B_y}&=\sum_{a,b=\pm1} a\,b\, P(ab|xy),\\
	\braket{A_x }&=\sum_{a=\pm1} a\, P(a|x).
\end{split}
\end{equation}

Specifically, the above inequality can be maximally violated by measuring the following  local observables
\begin{gather*}
A_0 = B_0=C_0=D_0=- \sigma_z,\\
A_1 = B_1=C_1=D_1= \sigma_x,\nonumber 
\end{gather*}
on the four-qubit state:
\begin{equation}
	\ket{\psi}=\frac{1}{2\sqrt{10-4\sqrt{5}}}\left[ (2-\sqrt{5})\ket{0001}+\ket{1110}+\circlearrowright\right],
\end{equation}
giving a value of 11+$8\sqrt{5}\approx 28.885$ for the left-hand-side of inequality~\eqref{Ineq_OptimalVis}. Equivalently, in terms of the {\em state visibility} $w$,\footnote{For rank-one projective measurements, the state visibility $w$ coincides with the visibility $v$ discussed in Sec.~\ref{Sec:MixPart} but in general, these values may be different. See, for instance, Ref.~\cite{MerminCGLMP}.} the mixture
\begin{equation}\label{Eq_state_visibility}
	w\ket{\psi}\bra{\psi}+(1-w)\frac{\one}{16}
\end{equation}
with $\one$ being the identity operator, is able to violate inequality~\eqref{Ineq_OptimalVis} for $w>\frac{19}{11+8\sqrt{5}}\approx0.6577$. Clearly, this improves over the best known visibility of $\tfrac{1}{\sqrt{2}}$ required to detect three-way NS-nonlocality (see, for instance, Refs.~\cite{Bancal:PRA:014102,BBGL}).

Let us also remark that, as one can verify using the techniques presented in Ref.~\cite{Liang:PRA:2007}, inequality~\eqref{Ineq_OptimalVis} cannot be violated by the four-partite GHZ state:
\begin{equation}
	\ket{\rm GHZ_4} = \frac{1}{\sqrt{2}}  (\ket{0000}+\ket{1111}).
\end{equation}
In fact, heuristic optimization results suggest that there are 3555 other families of symmetrical $\NS{2/2}$ inequalities that cannot be violated by $\ket{\rm GHZ_4}$. Most of these inequalities thus serve as candidates for certifying in a device-independent manner that certain correlations do not arise from $\ket{\rm GHZ_4}$ (see Ref.~\cite{sharam}).  Nevertheless, as one would expect, this highly-entangled quantum state does violate some of the symmetrical $\NS{2/2}$ inequalities. Indeed, it violates 3 families of the symmetrical $\NS{2/2}$ inequalities maximally. Among all the symmetrical $\NS{2/2}$ inequalities, the best state visibility that we found for $\ket{\rm GHZ_4}$ is $1/\sqrt{2}$, whereas 
for the four-partite $W$-state, 
\begin{equation}\label{Eq_WState4}
	\ket{W_4} = \frac{1}{2} (\ket{1000}+\ket{0100}+\ket{0010}+\ket{0001}),
\end{equation}
it is 0.7703. Heuristic optimization results suggest that there are in total 7645 families of symmetrical $\NS{2/2}$ inequalities that cannot be violated by $\ket{W_4}$, among which 1371 families also cannot be violated by $\ket{\rm GHZ_4}$. Otherwise, we see that these quantum states serve as a resource that cannot be reproduced by any of these biseparable but post-quantum resources.

\subsection{A Nontrivial Facet-defining Four-way NS-nonlocality Witness}

As mentioned above, not all the  symmetrical $\NS{2/2}$ inequalities are violated by extreme points of the $\NS{3/1}$ polytope. Among those that are not, only the following inequality 
\begin{align}
	I_{\rm NS_3} &=  \braket{A_0 B_0} + \braket{A_1 B_1} + 2  \braket{A_0 B_0 C_0 D_0} - \braket{A_1 B_0 C_0 D_0} \nonumber\\
	 &\quad- \braket{A_1 B_1 C_0 D_0}+ \braket{A_1 B_1 C_1 D_0}+ 2 \braket{A_1 B_1 C_1 D_1}\nonumber\\
	 &\quad\,+\circlearrowright\quad \stackrel{\NS{2/2}}{\le} 10, \label{Ineq_common}
\end{align}
can be violated by quantum correlation. In particular, one can verify using the hierarchy of semidefinite programs~\cite{Vandenberghe:SR:1996} due to Navscu\'es-Pironio-Ac\'in~\cite{QMP.Hierarchy1} (see also Refs.~\cite{QMP.Hierarchy2,Moroder:PRL:PPT}) that   --- in contrast with inequality~\eqref{Ineq_OptimalVis} --- this inequality is violated maximally by $\ket{\rm GHZ_4}$, giving a value of 12.8062 for the left-hand-side of Eq.~\eqref{Ineq_common}, corresponding to a state visibility of 0.7809.

Incidentally, this inequality also corresponds to a facet of  the $\NS{3/1}$ polytope. Hence, inequality~\eqref{Ineq_common}   represents a facet-defining inequality for the polytope formed by the convex hull of $\NS{2/2}$ and $\NS{3/1}$. In other words, a violation of this inequality immediately implies not only Bell-nonlocality, but also four-way NS-nonlocality.

\subsection{Device-independent Entanglement Witnesses for Genuine 4-partite Entanglement}

Using the upper bound technique from Ref.~\cite{Moroder:PRL:PPT} (see also Ref.~\cite{DIEW}), we managed to identify 37 inequivalent families of symmetrical $\NS{2/2}$ inequalities that are necessarily satisfied also by all correlations in $\Q_{3/1}$.\footnote{This was achieved by solving some appropriate semidefinite programs~\cite{Vandenberghe:SR:1996} with the help of YALMIP~\cite{YALMIP} to  a numerical precision of $10^{-6}$.The technique from Refs.~\cite{Moroder:PRL:PPT,DIEW} only provides a hierarchy of upper bounds on the all the $\NS{2/2}$ Bell expressions achievable by correlation in $\Q_{3/1}$. Thus, within the finite resource that we have, there is no guarantee that these bounds are tight and thus there may be more than 37 families of symmetrical $\NS{2/2}$ inequalities that cannot be violated by $\Q_{3/1}$ correlations.}  Moreover, 27 out of these 37 families can actually be violated by 4-qubit entangled quantum states. 
These include inequality~\eqref{Ineq_common} and 26 others which can also be violated by $\NS{3/1}$-local correlations. All in all, we are thus left with 26 non-trivial inequivalent families of device-independent entanglement witness (DIEW) for genuine four-partite entanglement; violation of these Bell-like inequalities do not imply the presence of genuine multipartite nonlocality --- not even the weakest form of {\em four-way} NS-nonlocality --- but still certifies the presence of genuine four-partite entanglement. Somewhat surprisingly, the best state visibility that one can get from all these 26 DIEWs is still marginally worse than that given by 
inequality~\eqref{Ineq_common}, which detects not only genuine four-partite entanglement, but also four-way NS-nonlocality.

\section{Concluding Remarks}
\label{Sec_Conclusion}

The resource aspect of quantum entanglement is always of interest in quantum information science. The recent progress in experimental control of multipartite systems (see, for instance, Refs.~\cite{Experiment3,Multi-exp}), in addition, provided strong motivation for one to understand better the resource aspect of multipartite quantum correlations. Using the definitions provided in Ref.~\cite{Bancal:PRA:014102}, we have taken a step in this direction by comparing the extent to which certain tripartite and four-partite quantum correlations can be reproduced by different kinds of biseparable correlations, in particular those whereby parties within the same subgroup are allowed to shared either non-signaling (NS) or one-way signaling resources. In the tripartite scenario, we have found that for certain quantum correlations arising from $W$-state, reproducing them using the above-mentioned non-quantum resource requires a mixture of  different groupings and/or time orderings of measurements. This means that within the framework allowed by these non-quantum resources, {\em mixtures} must also be seen as a resource in order to reproduce quantum correlations. The necessity to mix different time orderings also suggest that when seen from this classical perspective, these quantum correlations do not admit a definite causal order, cf. Refs.~\cite{Oreshkov:NatComm:1092,Bancal:NatPhys,Barnea:2013}.

Moving to the four-partite scenario where no complete characterization in terms of Bell-like inequalities is known, we managed to solve the symmetrized $\NS{2/2}$ polytope (with two binary measurements per party) and obtained 23,306 inequivalent families of Bell-like inequalities. Since this polytope is a superset {\em approximation} of the set of biseparable correlations with each group sharing at most a two-party NS resource, a violation of any of these inequalities immediately implies three-way NS-nonlocality in a four-partite scenario. Interestingly, there is even one family among all these whose violation also implies four-way NS-nonlocality and which can be violated by quantum correlations. 

The most robust quantum violation of these inequalities can tolerate 34.23\% of white noise and this offers a small advantage of 4.94\% when compared with that achievable in the  tripartite scenario having the same number of measurement settings and outcomes. Somewhat surprisingly, this inequality can provably not be violated by the four-partite GHZ state, which is otherwise known to give maximal violation of all full-correlation Bell inequalities in this scenario~\cite{WW:Complete:npartiteBI}. Among all these facet-defining inequalities for the symmetrized $\NS{2/2}$ polytope, we have also identified 26 families which can be used directly as device-independent entanglement witness whose violation implies genuine four-partite entanglement but not any form of four-way nonlocality.

Clearly, our work only represents a tip of the iceberg of what one can learn about the resource nature of multipartite quantum nonlocality. For instance, it will be desirable to perform similar characterization like the kind that we have achieved here in the four-partite scenario for one-way signaling resources, as this superficially resembles the scenario appearing in measurement based quantum computation~\cite{MBQC}. Research along this line may shed further insight on the connection between multipartite quantum nonlocality and its role in quantum computation~\cite{Anders:PRL:2009}.

On the other hand, we note that the huge list of symmetrical $\NS{2/2}$ inequalities obtained have given us the natural tools to investigate  {\em delocalized tripartite nonlocality}~\cite{DNL_def}, namely the phenomenon where three-way (i.e., tripartite) nonlocality is present in an $N>3$-partite scenario, but where this nonlocal feature cannot be attributed to any of the parties. A detailed analysis of this phenomenon using the tools developed here will be presented elsewhere~\cite{DNL_def}.

\begin{acknowledgments}
We acknowledge useful discussion with Jean-Daniel Bancal, Stefano Pironio and Nicolas Brunner. We also acknowledge assistance from Jean-Daniel Bancal and Tomer Barnea in solving the symmetrical $\NS{2/2}$ polytope. This work is supported by the UK EPSRC, the Swiss NCCR ``Quantum Science and Technology",  the CHIST-ERA DIQIP, and the European ERC-AG QORE.
\end{acknowledgments}

\end{document}